# Collective emotions online and their influence on community life

Anna Chmiel[1], Julian Sienkiewicz[1], Mike Thelwall[2], Georgios Paltoglou[2], Kevan Buckley[2], Arvid Kappas[3] & Janusz A. Hołyst[1*],

*1 Faculty of Physics, Center of Excellence for Complex Systems Research, Warsaw University of Technology, Warsaw, Poland*

*2 Statistical Cybermetrics Research Group, School of Technology, University of Wolverhampton, Wolverhampton, United Kingdom*

*3 School of Humanities and Social Sciences, Jacobs University Bremen, Bremen, Germany*

 * corresponding author: jholyst@if.pw.edu.pl

# Abstract

**Background:** E-communities, social groups interacting online, have recently become an object of interdisciplinary research. As with face-to-face meetings, Internet exchanges may not only include factual information but also emotional information – how participants feel about the subject discussed or other group members. Emotions in turn are known to be important in affecting interaction partners in offline communication in many ways. Could emotions in Internet exchanges affect others and systematically influence quantitative and qualitative aspects of the trajectory of e-communities? The development of automatic sentiment analysis has made large scale emotion detection and analysis possible using text messages collected from the web. However, it is not clear if emotions in e-communities primarily derive from individual group members' personalities or if they result from intra-group interactions, and whether they influence group activities.

**Methodology/Principal Findings:** Here, for the first time, we show the collective character of affective phenomena on a large scale as observed in four million posts downloaded from Blogs, Digg and BBC forums. To test whether the emotions of a community member may influence the emotions of others, posts were grouped into clusters of messages with similar emotional valences. The frequency of long clusters was much higher than it would be if emotions occurred at random. Distributions for cluster lengths can be explained by preferential processes because conditional probabilities for consecutive messages grow as a power law with cluster length. For BBC forum threads, average discussion lengths were higher for larger values of absolute average emotional valence in the first ten comments and the average amount of emotion in messages fell during discussions.

**Conclusions/Significance:** Overall, our results prove that collective emotional states can be created and modulated via Internet communication and that emotional expressiveness is the fuel that sustains some e-communities.

# Introduction

The scientific study of emotions began with the publication of Darwin's "The expression of the emotions in man and animals" in 1872 [1]. Based on this biological framework, *psychologists* have researched affective processes with regard to a) causes, b) mental processes and bodily systems involved, c) intra- and interpersonal regulation, d) communication, and e) consequences. For Darwin, the social nature of emotion was evident but since the end of the 19th century, particularly starting with William James [2], the majority of theories and research have focused on psychological processes within the individual, neglecting the complex behavior that emerges when individuals interact. In the past few years there has been increasing acceptance that the brain is a social organ [3] in the sense that the perception of others' emotional actions leads to complex behaviors that are still poorly understood. Nevertheless, insights have been gained regarding the five issues raised above, especially concerning social structures emerging due to emotional interactions.

The Internet can be studied as a system of human behavior in which social dynamics [4-12] are evident. Internet communication displays different activity patterns compared to traditional communication [13]. Many people spend increasing amounts of time online on social web sites (cyberspace) like MySpace, Facebook, Twitter, and a variety of blogs. Networks of people interacting in this way are often referred to as virtual communities, based on the influential book by Howard Rheingold (1994) [14]. While everyday connotations of the term community might trigger assumptions or connotations regarding the degree or quality of interaction of members, their social relationship, or the temporal and spatial properties of interactants that might not hold in a virtual community, the term has found common acceptance. Synonymous descriptors are online community or e-community.

Nowadays e-communities are increasingly prevalent and important in everyday life [15] as well as in business contexts where the use of social web functions plays an important part in marketing and customer interaction. Moreover, studies of affective interactions in e-communities are pivotal for understanding social relations in general as only online can large-scale interactions be studied systematically [4,9,12]. This paper uses automatic sentiment analysis to investigate such large scale Internet interactions to identify the key properties of emotion transmission in e-communities and to further our understanding of human interaction in larger social networks.

Although emotions are typically expressed using a variety of non-linguistic mechanisms, such as laughing, smiling, vocal intonation and facial expression, textual communication can be just as rich and has been augmented by expressive textual methods, such as emoticons and slang [16]. Taking advantage of this, sentiment analysis, a research field in computational linguistics and computer science, has evolved rapidly in the last ten years in response to a growing recognition of the importance of emotions in business and the increasing availability of masses of text in blogs and discussion forums. The development of a number of algorithms to detect positive and negative sentiment has also made large-scale online text sentiment research possible, such as predicting elections by analyzing sentiment in Twitter [17], and diagnosing trends for happiness in society via blogs [18] and Facebook status updates [19].

In this paper we discuss the impact of emotional expressions from Internet users on the vitality of online debates. We focus on (i) measuring the transfer of emotions between participants and (ii) the influence of emotions on a thread's life-span.

# Results

**Collective effects in emotional discussions** We collected over 4 million comments from three prominent interactive spaces: blogs, BBC discussion forums and the popular social news website Digg (for key properties see Table 1). The texts were processed using sentiment analysis classifiers to predict their emotional valence (see Figure 1 and Materials and Methods). To detect affective interactions between discussion participants we calculated statistics for groups of comments with similar emotion levels. Every discussion thread (identified by a unique URL) was analysed separately and was converted into a chain even if a tree structure was present (see Figure S1). We define an emotional cluster of size $n$ as a chain of $n$ consecutive messages with the same sentiment orientation: i.e. negative, positive or neutral, where before the cluster and after the cluster is a message with a valence different from the cluster valence (see the upper row in Figure 2 and Figure S2). For comparison we present also shuffled data received from the same discussions (see the bottom row in Figure 2). The clusters in the shuffled data are clearly shorter than the clusters in the original discussion. The reason for this could be emotional interactions between group members in the original data. To prove this hypothesis we checked the distribution of cluster lengths. If affective interactions between group members were absent then the probability of finding a cluster of length $n$ among all clusters corresponding to any specific emotion $e = \{-1, 0, 1\}$ (see Materials and Methods for details) would be described by an independent and identically distributed (i.i.d.) random process (see the second section in Text S1 of SI) with the cumulative distribution:

$$P^{(e)}_{i.i.d.}(\geq n) = p(e)^{n-1} \qquad (1).$$

Here $p(e)$ is the probability of a negative or positive emotion measured as the number of comments with the valence $e$ divided by the total number of comments in the considered data (for exact values see Table 1). Figures 3A and 3B show BBC and Digg data compared to predictions from an i.i.d. process. The agreement between the data and Eq. (1) diverges for $n > 10$, and the frequency of long clusters of the same emotional valence is large compared to the frequency expected for mutually independent messages. For example for BBC forums there were 91 negative clusters of length $n = 25$ while the i.i.d. prediction is $n = 6$; similarly for Digg there were 57 positive clusters of length $n = 11$ whereas the i.i.d. prediction is $n = 2$. It follows that there is a tendency for emotions of the same valence to occur together, suggesting the presence of attractive affective forces between discussion participants: posts tend to trigger follow-up posts of the same valence.

To quantify the strength of the interactions between authors of consecutive posts, consider the conditional probability $p(e|ne)$ that after $n$ comments with the same emotional valence the next comment will have the same valence. For an i.i.d. process such a conditional probability is independent from the parameter $n$ since by definition $p(e|ne) = p(e)$ for the i.i.d. process. Figure 3C shows however that this probability for the original data is an increasing function of $n$ for $n < 20$. The data reveals the relation

$$p(e|e) < p(e|ee) < ... < p(e|ne) \approx p(e|e)n^\alpha \qquad (2),$$

where $p(e|e)$ is the conditional probability that two consecutive messages have the same emotion defined by $p(e|e) = p(ee)/p(e)$ where $p(ee)$ is the joint probability of the pair $ee$ that is measured as a number of occurrences of the two consecutive messages with the same valence $e$ divided by the number all appearing pairs. The characteristic exponent $\alpha$ represents the strength of the preferential process leading to the *long –range* attraction between posts of the same emotion. When $\alpha = 0$ there is an i.i.d. process. Relation (2) implies that finding a positive message after seven positive comments is more likely than after six. It holds true for $n < 20$, but then saturation follows, with $p(e|ne)$ decreasing to zero for large $n$ (see Figure 3C). Preferential processes are common in complex systems [20] with positive feedback loop dynamics and they can be one of sources responsible for the emergence of fat-tailed distributions, including power-law scaling [21,22]. Using the relation $p(e|ne) \approx p(e|e)n^\alpha$ gives an analytical approximation to the cluster distribution (see the second section in Text S1 of SI).

$$P_\alpha^{(e)}(\geq n) \approx p(e|e)^{n-1}[(n-1)!]^\alpha \qquad (3).$$

This solution (3) is presented in Figure 3 (solid lines in Figures 3A and 3B). The fit with the data is far better than for the i.i.d., especially for large $n$. The differences between the analytical assumption and the real data come from the artificial extension of the scaling relation $p(e|ne)$, for large $n$ (see the second section in Text S1 of SI).

Values of the exponent $\alpha$ for different communities and different cluster types are presented in Figure 3D as a function of the probability $p(e)$ derived from the frequency of a given emotion. A good fit is $\alpha = 0.75\exp[-4p(e)]$ although power-law and linear approximations also work well, comparing the values of $R^2$ (see Figure description). This behaviour means that for more frequent emotions the chance to attach a consecutive message with the same valence grows slower with $n$ than for less frequent ones.

Figure 3 confirms that the occurrence of emotional posts cannot be described by the i.i.d. process and there are specific correlations between emotions in consecutive posts. These correlations result from emotional interactions between discussion participants via their messages. The interactions possess an attractive character because clusters of posts with the same emotional valence are longer than clusters from random distributions. The emotion expressed by a participant depends on the emotions in previous posts: he/she tends to express emotions that have been recently used in the discussion. This observation is consistent with general ideas regarding functions of emotions (e.g. [23]). Thus, positive emotions in interactive contexts could be interpreted as facilitating affiliative responses. With negative emotions the situation is more complicated as different emotional states, such as sadness and anger, would tend to elicit different specific responses in others. Anger, whether directly targeted at interactants or at the topic of the discussion, might elicit anger, sadness in contrast might elicit empathic responses that also expresses sadness. Given that current sentiment analysis algorithms cannot distinguish between such discrete states, a large-scale quantitative analysis of such hypotheses is, at present, not possible. However, interestingly enough the empirical data we present here might just suggest that it does not matter which discrete negative emotions are expressed - negative statements tend to follow negative statements.

**Life-spans of communities** Do community emotions evolve over time? This phenomenon was studied quantitatively for BBC forums as follows. Threads of the same size were grouped together and a moving average of the emotion type of the last 10 comments was calculated for each point. As seen in Figure 4A, shorter threads tend to start from a lower (i.e., less negative) emotional level than longer ones. On the other hand threads end with a similar mean emotional valence value regardless of their lengths: the last point of each data series in 4A (circles, squares, triangles and diamonds) is at almost the same level, about -0.42. This phenomenon is echoed in Figure 4B where the average emotional valence of the first 10 comments minus the average emotional valence of the last 10 comments is plotted, showing that longer threads have larger changes in emotional valence. Figure 4C also suggests that the initial emotional content (whether positive or negative) may be used as an indicator of the expected length of a thread: low absolute average emotion valences lead to shorter discussions. A possible heuristic explanation is that the first few posts in a thread may give it the potential (emotional fuel) to propel further discussion. Once the emotions driving the discussion dry out, the thread is no longer of interest to its participants and it may die. For threads possessing higher initial levels of emotion it takes more comments to resolve the emotional issue, resulting in longer discussions (cf. results for the discussions at BBC forum in [24]).

## Discussion

Understanding the nature of interactions in e-communities is useful for predicting the future shape of society because of the increasing importance of the Internet in communication [6-9]. Since this medium offers both anonymity and the possibility to contact with many others it is important to understand the impact of collective effects within online discussions. Phenomena such as an online disinhibition effect [25] might impact the intensity of emotions expressed which in turn might create differences in the contagion from and reactions to statements made in interaction.

Here, on the basis of automatic sentiment detection methods applied to huge datasets we have shown that Internet users' messages correlate at the simplest emotional level: positive, negative or neutral messages tend to provoke similar responses. This result agrees with observations of singular events corresponding to propagations of emotions in bipartite networks of blogs [8]. Our simple approach demonstrates that the existence of many groups of consecutive messages (i.e., clusters) with the same emotional valence can be explained by preferential processes for cluster growth. The collective character of expressed emotions occurs in several different types of e-community. It was observed for BBC forums and Digg, both communities where negative emotions dominate, and also for the Blogs06 blogs where most comments are positive. This may seem to contradict a previous analysis of online social networks [9] that found negative interactions to be different from positive interactions. However, the links and network motifs in the e-community studied in [9] were not expressing the emotions of participants but relationships like friendship, communication, enmity or punishment.

We hypothesise that the strength of emotional interactions can be indirectly measured by the parameter $\alpha$ expressing the influence of the most recent emotional cluster on the probability that the next post has the same emotion (Eq. (2)). Table 1 shows that this strength depends not only on the kind of e-community but also on the value of the emotional valence. Surprisingly, stronger collective behavior, corresponding to larger values of $\alpha$, exists when a given emotion is less frequent.

We are aware of the fact that both our method of data collection and the results of the sentiment classifiers suffer from various errors that can be only partly estimated (see Materials and Methods Section and Ref [26]).These errors, however, are unlikely to lead to a spurious occurrence of the observed clustering effects. Since the classifier treats every post independently from the previous one, there is no memory effect that could be introduced by classifier actions. Assuming that classifier errors are random, it is more likely that a cluster is broken by a random error than that a cluster is formed by a series of errors. Hence the raw cluster statistics are likely to underestimate the strength of the clustering phenomenon existing in the studied communities

We also give evidence that in BBC forums the initial emotional level of a discussion fuels its continuation: when this fuel is exhausted a discussion is likely to end (Figures 4A, B). This is because higher levels of emotions in the first ten comments in a thread lead to longer discussions (Figure 4C). Although this behavior agrees with observations concerning political discussions in Polish Internet forums [7] where *the growth of discussions was dependent on the degree of controversy of the subject and the intensity of personal conflict between the participants* [7] the effect was not present in the Digg and Blogs06 data so the type of e-community matters for this phenomenon.

Our analysis provides a better understanding of affective interactions between large numbers of people and is an important step towards the development of models of collective emotions in cyberspace [8,24,27-31]. By involving collective phenomena, these patterns go beyond the complexity of the nested intra- and inter-individual feedback loops in face-to-face communication [32]. In the future software tools [33-34] may be designed to support e-communities by measuring the emotional level of discussions. Since negative emotions not only prolong discussions but can also damage cooperation between community members, emotion level information may help participants to keep a community alive.

## Materials and Methods

**Data sets.** The BBC web site had a number of publicly-open moderated *Message Boards* covering a wide variety of topics that allow registered users to start their own discussions and post comments on existing discussions. Our data included discussions posted on the Religion and Ethics and World/UK News message boards starting from the launch of the website (July 2005 and June 2005 respectively) until June 2009. The Blogs dataset is a subset of the Blogs06 [35] collection of blog posts from 06/12/2005 to 21/02/2006. Only posts attracting more than 100 comments were extracted, as these seemed to initialize non-trivial discussions. The Digg dataset comprises a full crawl of digg.com, one of the most popular social news websites. The data spans February to April 2009 and consists of all the stories, comments and users that contributed to the site during this period [36].

**Algorithms.** Sentiment analysis algorithms typically operate in three stages: (a) separate objective from subjective texts, (b) predict the polarity of the subjective texts, and (c) detect the sentiment target [36]. A variety of methods are used, including machine learning based upon the words used in each text, summarized in vector form [37], and lexical approaches that start with a dictionary of known sentiment-bearing terms and apply linguistically-derived heuristics to predict polarity from their occurrence and contexts [38]. Our algorithm used supervised, machine-learning principles [39]. This is an efficient way to generate a classifier when there is a large amount of human-coded documents because the algorithm can use these to learn the features of documents that typically associate with the different categories. We implemented an hierarchical extension of a standard Language Model (LM) classifier [39]. LM classifiers estimate the probability that a document belongs to all of the available classes and select the one with the highest probability as the final prediction. In our hierarchical extension a document is initially classified by the algorithm as objective or subjective and in the latter case a second-stage classification determines its polarity, either positive or negative. We used a manually annotated subset of approximately 34,000 documents from the Blogs06 data set as a *training corpus*. After training, the algorithm would have learned which words typically occur in positive, negative and objective documents. For instance, it can be expected to learn that documents typically containing "love", "hate" or "disagree" are likely to be subjective and that "love" and "hate" are good predictors for the positive and negative categories respectively. Because the distribution of documents per category is uneven, the probability thresholds for both classification tasks were optimized on a small subset. The optimized classifier (described fully elsewhere [40]) has an accuracy of 73.73% for subjectivity detection and 80.92% for polarity detection on a humanly annotated BBC subset. A limitation of LM classifiers in contrast to linguistic algorithms is that they are context-insensitive and can make incorrect predictions when the normal meaning of individual words is changed by their context. The most common case is probably negations: a linguistic classifier would probably identify "not happy" as negative because "not" modifies the meaning of "happy" whereas a simple LM classifier is likely to ignore "not" altogether as a word that is neutral on its own and identify "happy" as a positive word, making this the polarity prediction. With a large training set, however, LM classifiers can learn the typical polarity of enough words to outperform the linguistic approach in some cases, including the data used here [40]. This may be partly due to the use of sentences and argument structures in our data set that are too complex for simple lexical rules and tend to neutralise their effectiveness. Note that other researchers have proposed many alternative LM variations to the fairly general approach used here. Our specific LM implementation is likely to vary in overall performance with these but the classifications should not change in a systematic way because the same data features are used as the input in all cases.

**Valence.** There is converging evidence that (hedonic) valence is at the center of emotion experience. This has been demonstrated using a variety of methods [41]. In simple terms, the relevant aspect of any object that elicits emotions is whether we like it or not, or whether it is good for us or not [42]. This distinction is evident from measures of approach/avoidance that clearly indicate that valence is one of the most important determinants of behavior from simple life-forms to humans [43]. Thus, while humans often report affect in terms of basic emotions, such as "happiness", "anger", or "sadness", the degree to which events or objects are evaluated as positive or negative that affects subjective experience (feeling), physiological processes in the periphery and the brain [44], expressive behavior, and action readiness [23]. For the purpose of the present model we shall identify valence as a single dimension that is scaled from -1 (negative) through 0 (neutral), to +1 (positive). There is a debate to whether it is possible that positive and negative emotions can co-exist [45] but for practical purposes a single dimension represents emotions well. Similarly, a second (arousal) and occasionally a third dimension (e.g., dominance or power) are considered in dimensional models (see [46]). However, valence explains most of the variance from these dimensions [47]. Thus, for simplicity in the present analysis we use only the primal dimension of affect – valence.

**Acknowledgements**
This work was supported by a European Union grant by the 7th Framework Programme, Theme 3: Science of complex systems for socially intelligent ICT. It is part of the CyberEmotions (*Collective Emotions in Cyberspace*) project (contract 231323). We are thankful to all members of CyberEmotions consortium for useful comments and discussions. JAH, ACh. and JS acknowledge support from Polish Ministry of Science Grant 1029/7.PR UE/2009/7 JAH, ACh and JS would like to thank Agnieszka Czaplicka, Piotr Pohorecki and Paweł Weroński for their help with data cleansing.

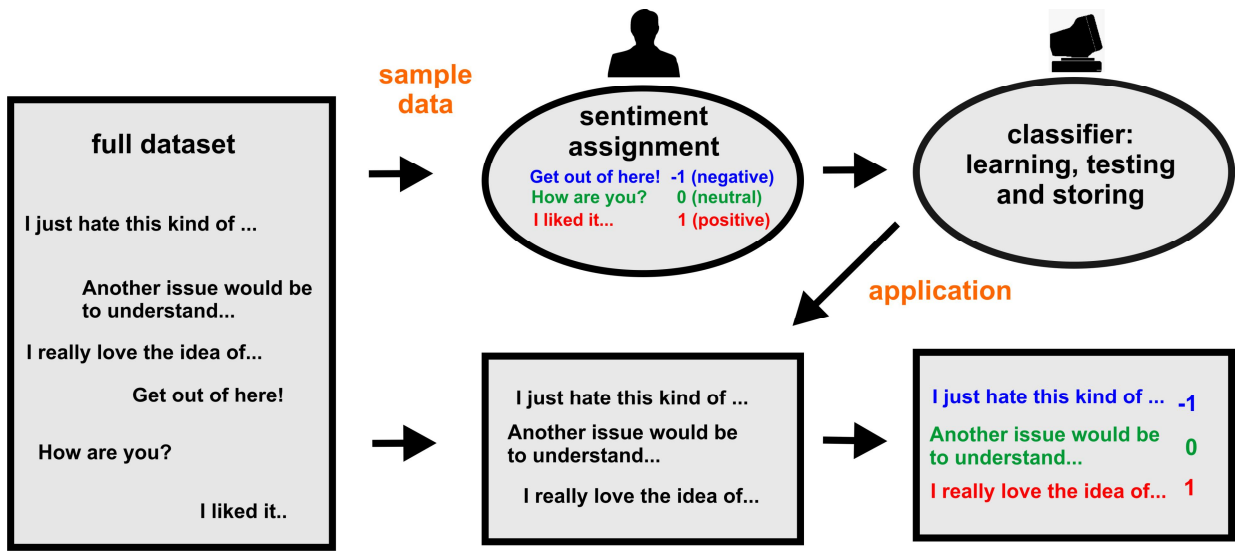

**Figure 1 Schematic plot of the process of document classification.** A sample from the set of documents is passed to human experts who read the content and manually classify it. The algorithm extracts the characteristics of each class by analyzing the provided documents, i.e., "learns by example", and stores this knowledge. As result each document is classified with the emotional value –1, 0 or +1 describing its emotional valence.

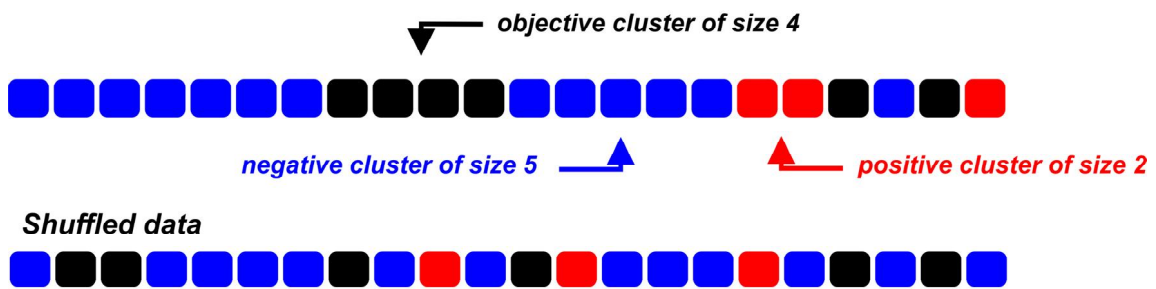

**Figure 2 An example of a discussion in the "Eastern religion" BBC forum in September 2005.** The original discussion, consisting of 22 posts is shown in the upper row. Each square represents one post: red, blue or black squares indicate that the comment was classified as, respectively, positive, negative or neutral (objective). The bottom row presents shuffled data, i.e., the comments were arranged in a random order. Note the difference between the length of clusters in the original and in the shuffled data.

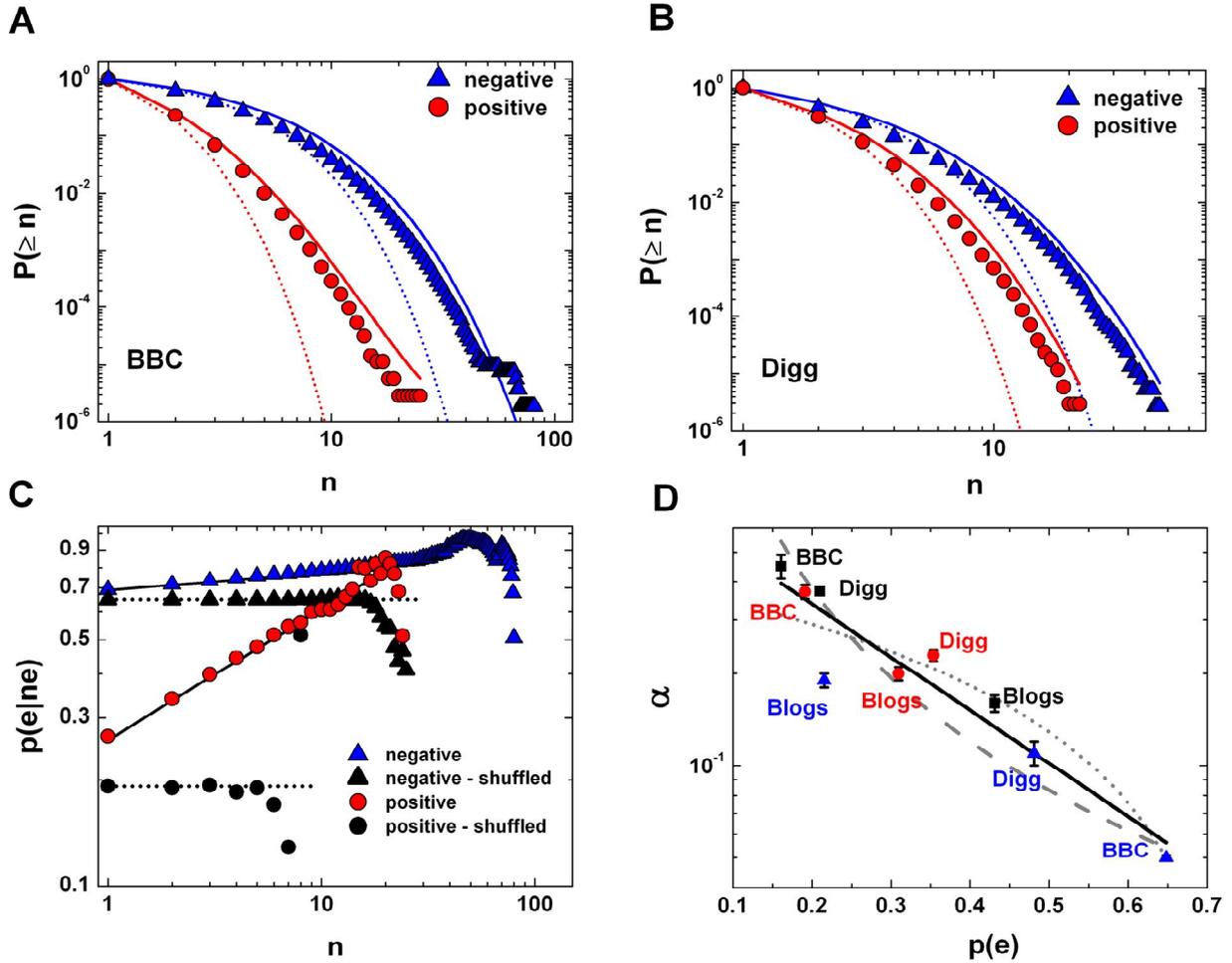

**Figure 3 Emotional clustering in Digg, BBC and Blog data.** (A-B) Cumulative cluster distributions for the BBC (A) and Digg data (B). Points correspond to the collected data (circles - positive messages, triangles - negative messages). Dotted lines are the i.i.d. process, and solid lines derive from Eq (3). It is clear that messages are mutually dependent and using the model with preferential cluster growth leads to a better fit with collected data. (C) The conditional probability of the next comment occurring having the same emotion for BBC (open symbols) data. Circles come from positive messages and triangles from negative messages. We observe that $p(e|ne) = p(e|e)n^{\alpha}$ (solid lines) in the first ten where there is the best statistics available (the largest number of comments) of the gathered data. Filled symbols indicate the shuffled data and dotted lines show the values of emotional probabilities $p(+)$ and $p(-)$ for BBC data (see Table 1). Note that $p_{i.i.d.}(e|ne) = p(e)$ fits well to shuffled data for small values of $n$. (D) The preferential exponent $\alpha$ decays with emotion frequency $p(e)$ although no exact relation can be received from the collected data. The solid line follows the relation $\alpha = B\exp[-\beta p(e)]$ with $B = 0.75 \pm 0.14$ and $\beta = 4.0 \pm 0.5$ while dotted and dashed curves are, respectively, power-law and linear fits. The value of $R^2$ for exponential fit is 0.96, while for power-law and linear it is 0.94 and 0.90 respectively.

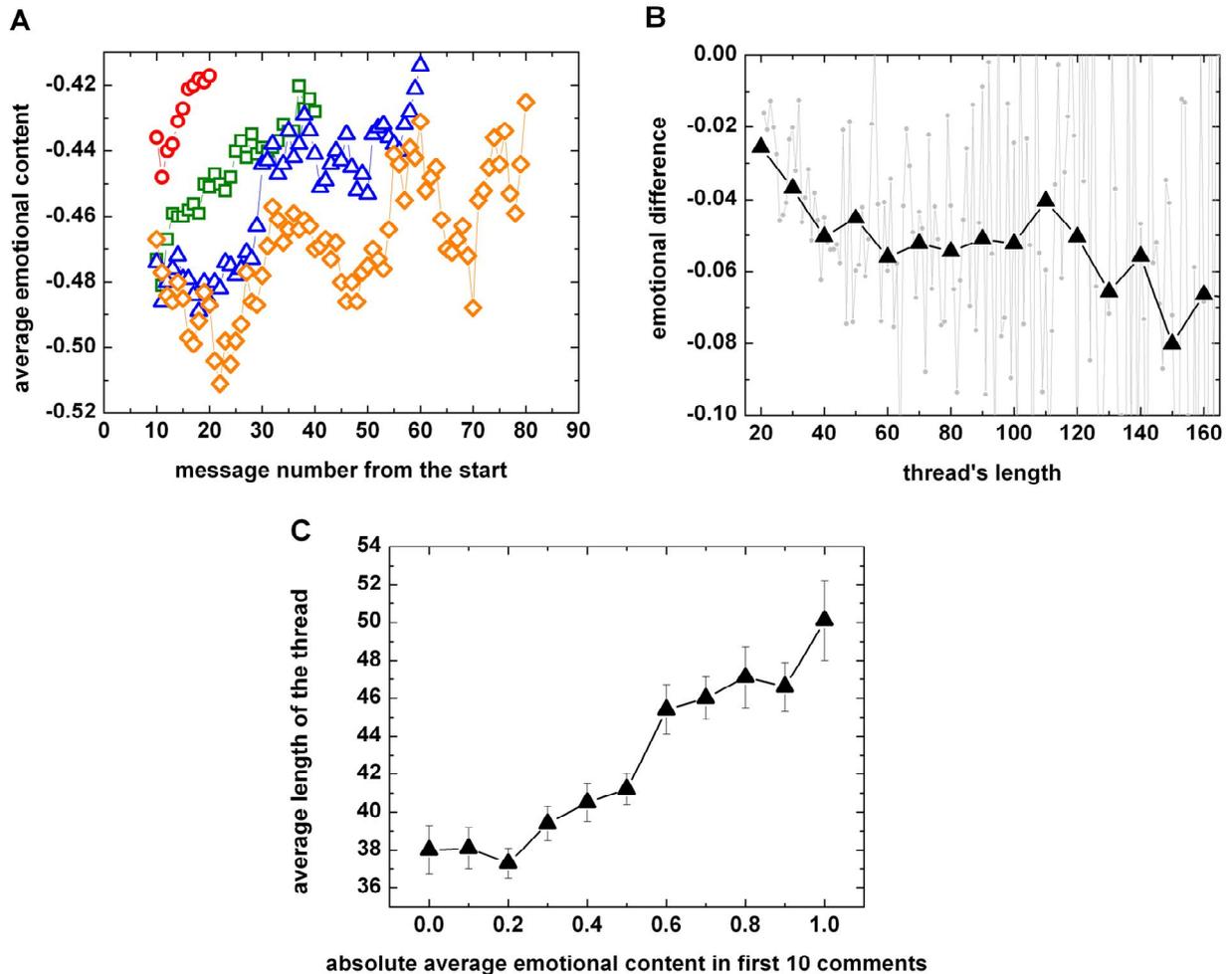

**Figure 4 Time dependence of emotions in BBC forum threads.** (A) Average emotion valence in the thread (moving average of the previous 10 messages in the thread). Four groups of threads of lengths 20, 40, 60 and 80 are represented by different symbols (respectively circles, squares, triangles and diamonds). Shorter threads start from emotional levels closer to zero. (B), Emotional level (valence) at the beginning of a thread minus the emotional level at the end as a function of thread length (grey symbols). Black triangles display binned data. Longer threads use more emotional 'fuel' over time. (C), Average length of the thread as a function of the absolute value of the average emotion valence of the first 10 comments. Emotional thread starts, whether positive or negative, usually lead to longer discussions. Error bars indicate standard deviations.

**Table 1. Datasets properties**

|       | $N$ | $U$ | $T$ | $\langle e \rangle$ | Probability of emotion $p(e)$ | | | Exponent $\alpha$ | | |
|-------|-----|-----|-----|------|------|------|------|------|------|------|
|       |     |     |     |      | $p(+)$ | $p(-)$ | $p(0)$ | $\alpha_+$ | $\alpha_-$ | $\alpha_0$ |
| **BBC**   | 2,474,781 | 18,045 | 97,946  | -0.44 | 0.19 | 0.65 | 0.16 | 0.38 | 0.05 | 0.45 |
| **Digg**  | 1,646,153 | 84,985 | 129,998 | -0.16 | 0.31 | 0.48 | 0.21 | 0.20 | 0.11 | 0.37 |
| **Blogs** | 242,057   | N/A    | 1219    | 0.14  | 0.35 | 0.22 | 0.43 | 0.23 | 0.19 | 0.16 |

Properties of the three datasets: number of comments $N$, number of different users giving comments $U$, number of discussions/threads $T$, average valence in the dataset $\langle e \rangle$, probability of finding positive, negative or neutral emotion (respectively $p(+)$, $p(-)$ and $p(0)$) and values of exponents $\alpha$ for positive, negative and neutral clusters (respectively $\alpha_+$, $\alpha_-$ and $\alpha_0$). In case of Blogs data it was impossible to quantify the number of different users and note also a low number of comments in this dataset. Each data set has a different overall average valence – BBC is strongly negative, Digg is mildly negative while Blogs are mildly positive.

# Collective Emotions Online and Their Influence on Community Life – Supporting Information File S1

*Anna Chmiel, Julian Sienkiewicz, Mike Thelwall, Georgios Paltoglou, Kevan Buckley, Arvid Kappas and Janusz A. Hołyst*

**S1 Data structure.** The three datasets on which sentiment analysis was performed are characterized by different structures. In the case of Blogs06, all posts were originally arranged in chains of successive comments, as shown in Fig. S1B. In other words new posts are automatically added after the last post. On the other hand, the BBC and Digg data are arranged in a forum-like structure (see Fig. S1A), meaning that each user may make a comment to any previous post, thus starting a separate discussion. However, for the purpose of this study each discussion (thread) in each dataset was arranged chronologically, as in Fig. S1B. In this way it was possible to compare these different communities. Although BBC and Digg have a forum-like structure, the default view presented to the user was chronological. Thus a chronological simplification for analysis can be justified.

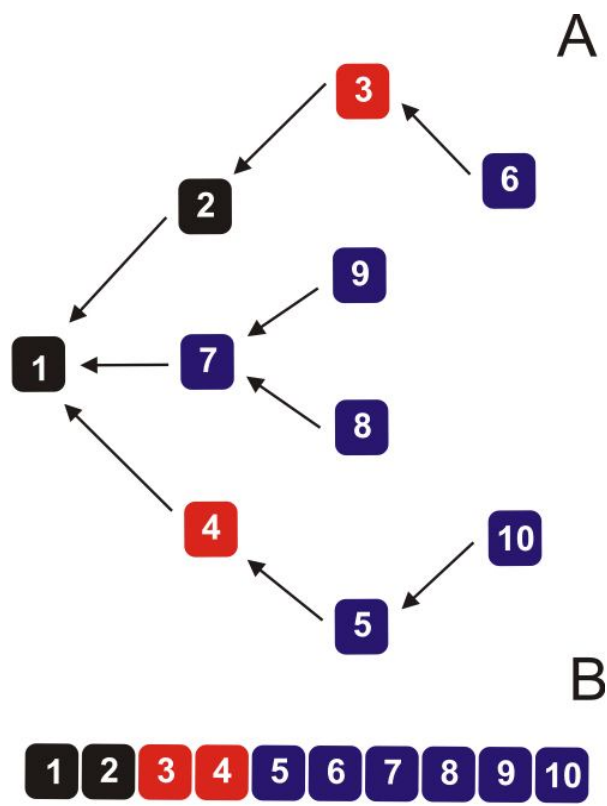

**Fig. S1.** The difference between the actual tree structure (A) present in the BBC and Digg datasets as compared to the chronological layout of the posts (B). The numbers indicate the order of messages (1 being the first, 10 being the last) while arrows indicate that a post was given in reply to another one (e.g. post 9 is the response to post 7).

**S2 Emotional cluster distributions.** An independent and identically distributed (i.i.d.) random process [1] corresponds to the simplest stochastic process where there is no statistical dependence between events at consecutive time-steps and at every time-step the event probability distribution is the same. When the parameter *p(e)* describes the probability of negative, positive or neutral emotion then the probability to find a cluster of size *n* among all clusters of the same valence *e* (see Fig. 2) scales as *[1-p(e)]²p(e)ⁿ*. Here the factor *[1-p(e)][1-p(e)]* corresponds to an event other than *e* just before and just after the cluster, i.e., where a change in valence takes place (for details see Fig. S3). Taking into account the normalization $\sum_{n=1}^{n=\infty} P_{i.i.d.}^{(e)}(n) = 1$ condition we have

$$A \sum_{n=1}^{n=\infty} [1-p(e)]^2 p(e)^n = A[1-p(e)]p(e) = 1 \quad (S1),$$

where *A* is the normalization constant (the sum starts from *n=1* as we do not take into account clusters of size 0). Using simple algebra we obtain the cluster distribution $P_{i.i.d.}^{(e)}(n) = $ *[1-p(e)]p(e)ⁿ⁻¹* and the resulting cumulative distribution is

$$P_{i.i.d.}^{(e)}(\geq n) = p(e)^{n-1} \quad (S2),$$

as represented by a dotted line in Fig. S2.

A Markov chain [2] is a basic stochastic process with one memory step when the probability of the next time state depends only on the previous one by corresponding conditional probabilities. The probability of finding a cluster of size *n* scales as *[1-p(e)][1-p(e|e)]p(e)p(e|e)ⁿ⁻¹* and, similarly to the i.i.d. process, the factor *1-p(e)* corresponds to any event just before the cluster other than *e* and the factor *1-p(e|e)* to the event just after the cluster. Taking into account the normalization condition (following the same scheme as in the i.i.d. case) we have the cluster distribution $P_M^{(e)}(n)=$ *[1-p(e|e)]p(e|e)ⁿ⁻¹* . Finally the cumulative distribution is

$$P_M^{(e)}(\geq n) = p(e|e)^{n-1} \quad (S3),$$

as represented by a dashed line in Fig. S2.

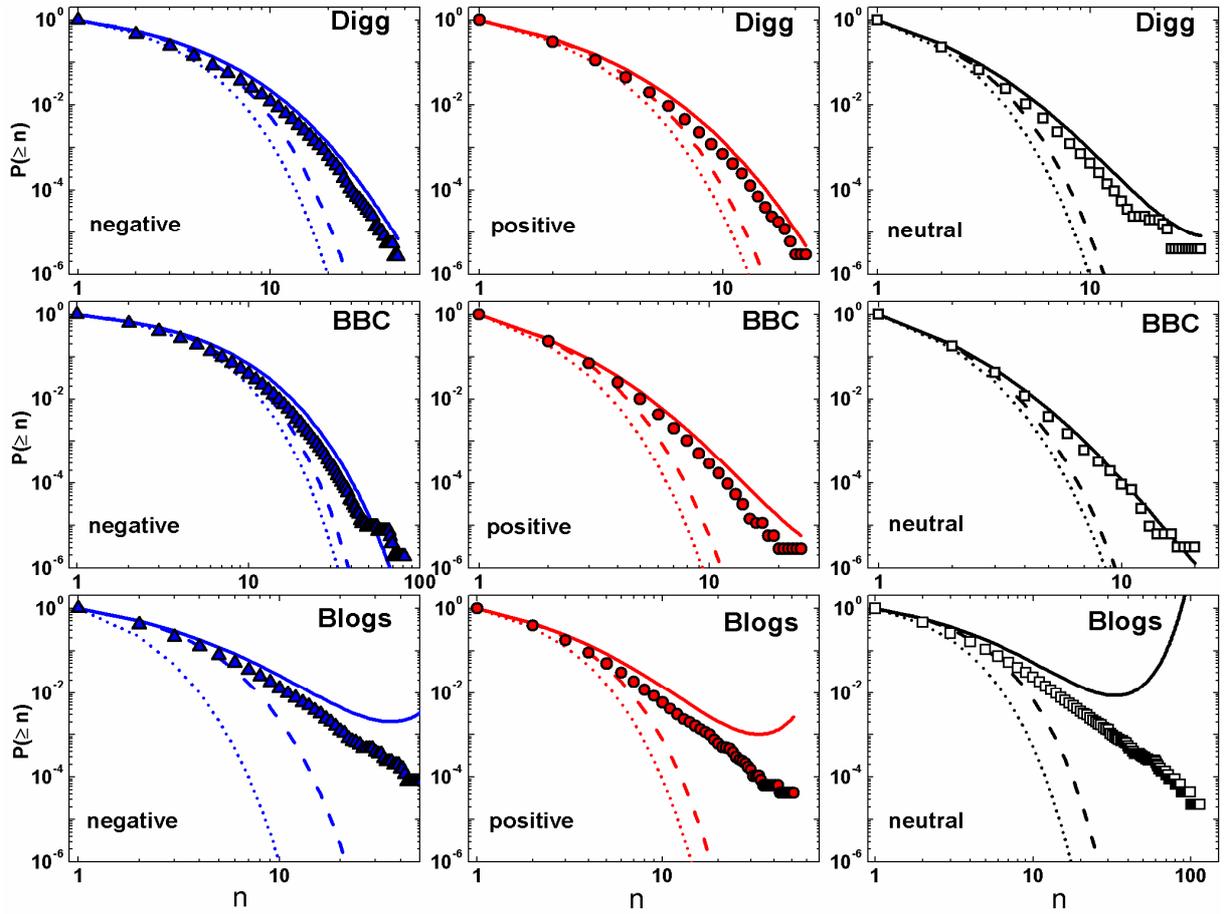

**Fig. S2.** Cumulative distribution of the cluster size for all data used in the study. Symbols are data (blue triangles, red circles and white squares, respectively for negative, positive and neutral clusters), dotted lines are i.i.d. processes given by Eq. (S2), dashed lines are Markov processes given by Eq. (S3) while solid lines come from Eq. (S6) and represent distributions based on the preferential attraction rule. The spurious increase of $P^{(e)}_\alpha$ ($\geq n$) for $n \geq 40$ for Blogs data is due to violation of the scaling (S4).

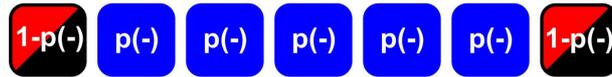

**Fig. S3.** In case of the i.i.d. random process, to obtain the probability of finding a cluster of <u>exactly</u> $n$ consecutive emotional values (here $n=5$ and $e=-1$) one has to take into account two factors: the length of the cluster itself and the issue that on the both borders there should be posts with emotional values other than inside the cluster. Thus, in the presented case, the probability is proportional to *[1-p(-)]p(-)$^5$[1-p(-)]*.

The conditional probability *p(e|ne)* that expresses the chance that after *n* messages with the same emotional valence *e* the next post will continue this trend is shown in Fig. S4 as symbols. The phenomenon of preferential attraction

$$p(e \mid ne) = p(e \mid e) n^\alpha \qquad (S4)$$

is evident for all datasets. The range of this scaling varies for different e-communities and different emotional valences of clusters – e.g., for the BBC neutral clusters we find a good fit for the whole range of the data. Note that the scaling can not be valid for very large n since $p(e|ne)$ must be smaller than 1. Preferential attraction plays a crucial role for cluster distribution. An approximation to real data behavior can be obtained by extending the relation (S4) up to the maximal cluster size in the considered community. The resulting cluster probability distribution for clusters of the size n scales as $[1-p(e|e)n^\alpha][p(e|e)]^{n-1}[(n-1)!]^\alpha$ where the first factor corresponds to an event other than $e$ just after the cluster of size $n$. This result resembles the previous Markov formula with additional factors reflecting the preferential effect. The analytical form of the normalization factor can be obtained only as an approximation when

$$p(e|e)^{n_{max}} (n_{max}!)^\alpha \approx 0 \qquad (S5)$$

The relation (S5) is fulfilled provided that $n_{max}$ is not too large ($n_{max} < 40$) or $\alpha$ is small ($\alpha < 0.1$). These conditions hold for all of the data sets except for the neutral clusters in Blogs06. As a result, the cumulative cluster distribution is

$$P_\alpha^{(e)}(\geq n) \approx p(e|e)^{n-1}[(n-1)!]^\alpha \qquad (S6)$$

This approximation is presented in Fig. S2 with solid lines. The fit to the data is far better than in the case of the i.i.d. assumption or even than the Markov approach, especially for large $n$. The observed differences between (S6) and the real data come from the artificial extension of the scaling relation (S4). It leads to the spurious behavior of $P^{(e)}_\alpha (\geq n)$ when it increases for large n (see Blogs06 clusters in Fig. S2).

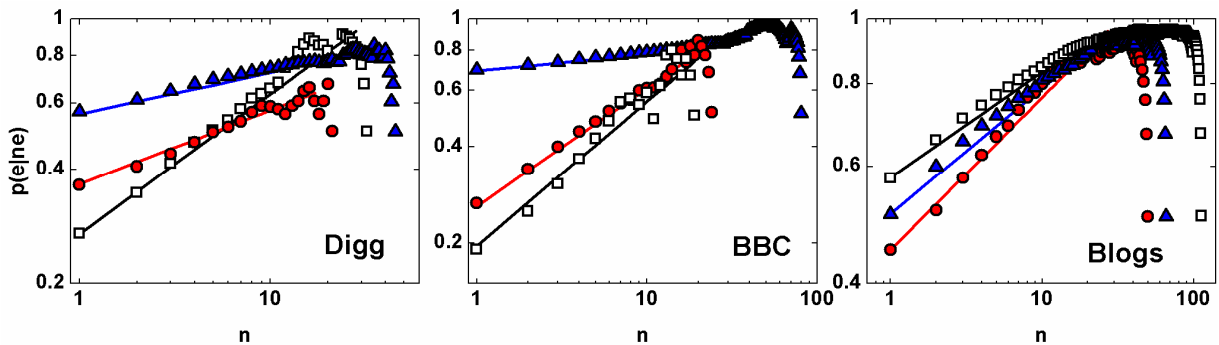

**Fig. S4.** The conditional probability $p(e|ne)$ of the next comment occurring having the same emotion for Digg, BBC and Blogs06 data. Symbols are data (blue triangles, red circles and white squares, respectively for negative, positive and neutral clusters) and lines reflect the fit to the preferential attraction relation (S4).

**S3 Conditional Probability.** In Fig. S5 we investigate more carefully the power-law character of the conditional probability *p(e|ne)*. We are aware of the fact that the regime of power law scaling is very short, although a comparison between power-law and linear fits, shows that the first function is in much better agreement with the data. Especially for the first point we can observe a large divergence between those fits, which is significant taking into account the fact that the first point has the best statistical quality (i.e., the number of events at *n=1* is much larger than for *n>10*).

|        | Digg |      |      | BBC  |      |      | Blogs |      |      |
|--------|------|------|------|------|------|------|-------|------|------|
| e:     | +    | 0    | -    | +    | 0    | -    | +     | 0    | -    |
| p(+|e) | 0.37 | 0.30 | 0.27 | 0.27 | 0.19 | 0.14 | 0.45  | 0.32 | 0.27 |
| p(0|e) | 0.20 | 0.27 | 0.17 | 0.15 | 0.20 | 0.17 | 0.39  | 0.58 | 0.22 |
| p(-|e) | 0.43 | 0.42 | 0.56 | 0.58 | 0.61 | 0.69 | 0.16  | 0.10 | 0.51 |

**Table S1. The conditional probabilities for all datasets** Each cell gives a conditional probability $p(e_2|e_1)$ of two post with emotions $e_1$ and $e_2$. For example for Digg data *p(+|0)=0.30*.

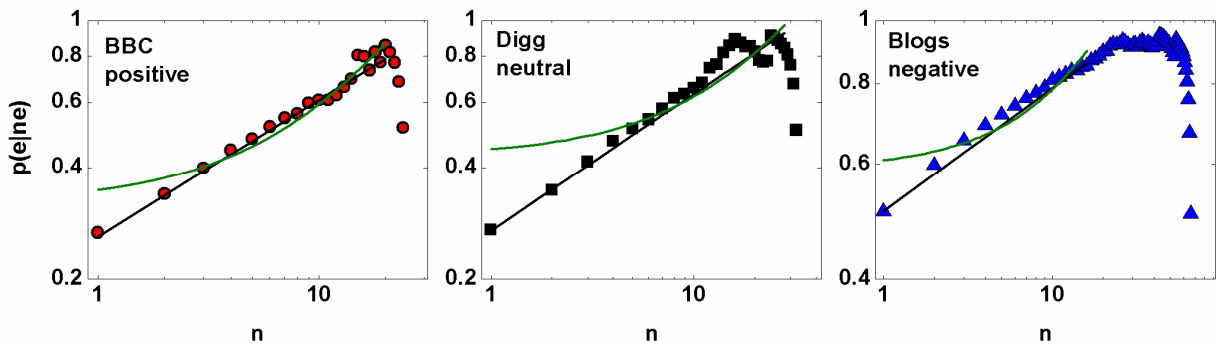

**Fig. S5.** A comparison between the power-law and linear fits to the conditional probability *p(e|ne)* for Digg neutral, BBC positive and Blogs06 negative posts. Symbols are data (blue triangles, red circles and black squares, respectively for negative, positive and neutral clusters), black lines reflect the fit to the preferential attraction relation (S4) and green lines are linear fits in the same regime.

**S4 Influence of the thread length on the emotional cluster distributions.** In all three cases the discussion lengths vary considerably. Figure S6 presents thread length distributions for BBC (left), Blogs (center) and Digg (right) data. The majority of discussions are shorter than 10 comments (this in not the case for Blogs where there are only threads with at least 100 comments – see the *Data sets* paragraph in the *Materials and Methods* section in the main text), yet there is also a significant number of very long threads in the fat-tail part of all three distributions. However, it is essential to notice that the thread length $L$ is not directly connected to the maximal cluster size observed in the data. Figure S7 presents a comparison between the BBC cumulative cluster distribution calculated for all data and subsets characterized with different thread lengths ($L=20$, $L=50$ and $L=100$). As one can see, for the negative posts, the maximal cluster size in threads with $L=50$ is even longer than in the case of $L=100$.

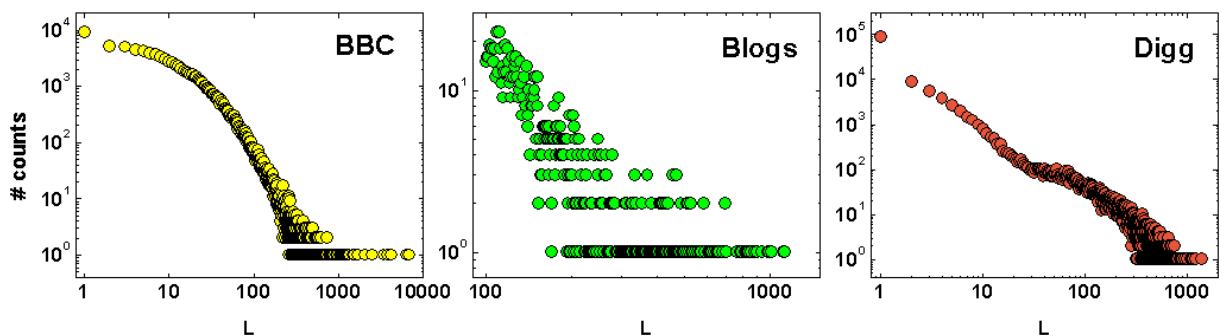

**Fig. S6.** Thread length distributions of the BBC (left) [3], Blogs (center) and Digg (right) data.

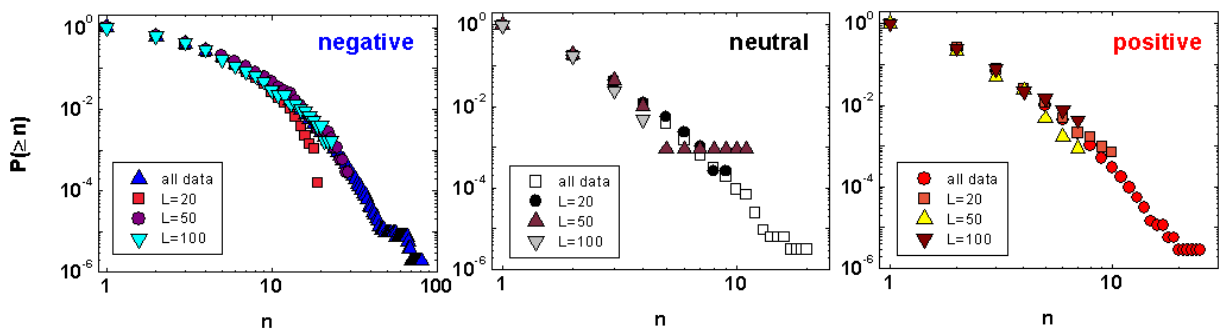

**Fig. S7.** Cumulative cluster distribution of the BBC negative (left), neutral (center) and positive (right) data for different subsets of thread lengths $L$.

**S5 Unique users in clusters.** To prove the validity of the main aspect of the paper – the transmission of emotions between discussion participants – it is crucial to check if the long threads are not dominated by a single user or just two people. Figure S8 shows the average number of unique users $<U>_n$ (i.e., users with a unique id) that take part in emotional clusters of size $n$. It suggests that although the longer discussions are dominated by a limited number of unique users, the majority of clusters is still created by a significant number of unique users (e.g., for negative clusters of size *n*=10 there are an average of 6 unique users).

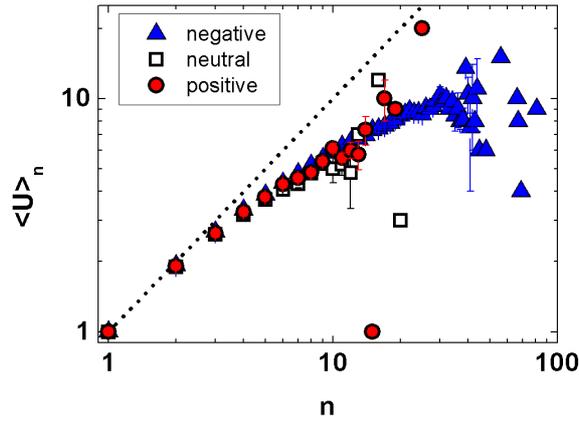

**Fig. S8.** Average number of unique users in the cluster $<U>_n$ versus the size of the cluster for BBC negative (triangles), neutral (squares) and positive (circles) data. The dotted line marks the relation $<U>_n = n$.

**S6 Comparing the cluster distribution with the k-1 Markov model.** Brendel at al. [4] used a *k-1* Markov model where the occurrences of a specific *k*-element sequence of nucleotides were obtained with information from the *k-1* previous events. The difference between the expected number of sequences and the observed value shows that, in the case of nucleotides, a Markov chain of order less than 4 does not give a good agreement.

To test how a *k-1* Markov model can be applied to the emotional statistics we defined the normalized difference *std(n)* between the expected and observed number of sequences with uniform emotional valence (an *n*-sequence is a chain of *n* messages with the same emotional valence; it is not the same as an *n*-cluster because, for example, sequences of size 4 can be part of a cluster of size 10):

$$std(n) = \frac{(Ns(n) - Es(n))}{\max\{\sqrt{Es(n)}, 1\}} \tag{S7}$$

Where the *Ns(n)* is the number of observed sequences of size *n*, and *Es(n)* is the expected number of size *n* obtained according the equation:

$$Es(n) = Ns(n-1) p(e \mid n-2). \tag{S8}$$

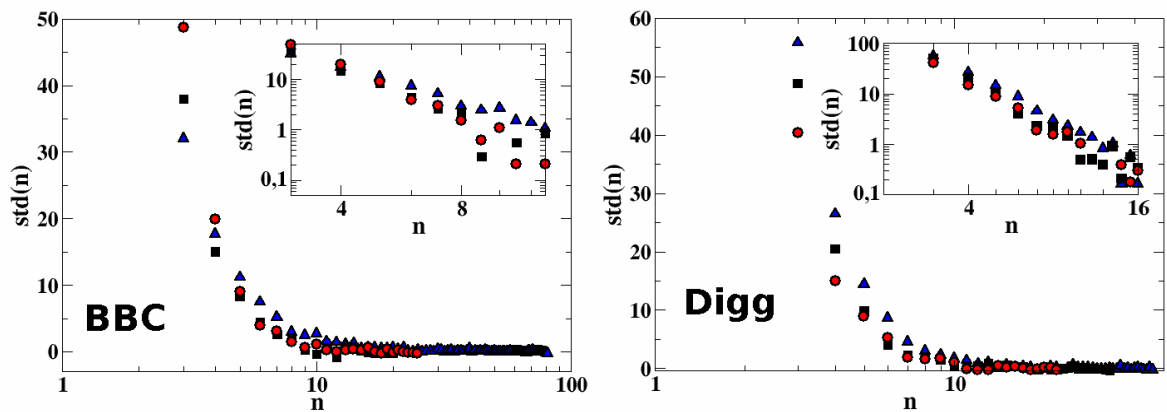

**Fig. S9** The normalized difference *std(n)* (S7) between the observed number of sequences of size *n* and the number expected by the *k-1* Markov model versus the size of the sequence *n* for BBC(left) and Digg (right). Blue triangles show negative sequences, red circles represent positive sequence and the black squares are neutral sequences. The insets shown in a log-log scale focus on the first part of relation where *std(n) >0* and the power law relation between *std(n)* and *n* can be observed. Only sequences larger than 3 can be handled with this method [5].

The results, presented in Fig S9, allow us to conclude that in the case of emotional sequences a Markov model of order less than 10 does not perfectly fit the data. This is another argument proving that a long rage correlation exists in building the sequences. Moreover, it should be underlined that the *k-1* Markov model uses almost full information from the data, contrary to the method presented in our paper.